# AUTOMATIC CONVERSION OF RELATIONAL DATABASES INTO ONTOLOGIES: A COMPARATIVE ANALYSIS OF PROTÉGÉ PLUG-INS PERFORMANCES

Kgotatso Desmond Mogotlane[1] and Jean Vincent Fonou-Dombeu[2]

[1]The South African Mint Company, Pretoria, South Africa
[2]Department of Software Studies, Vaal University of Technology, Vanderbijlpark, South Africa

## ABSTRACT

*Constructing ontologies from relational databases is an active research topic in the Semantic Web domain. While conceptual mapping rules/principles of relational databases and ontology structures are being proposed, several software modules or plug-ins are being developed to enable the automatic conversion of relational databases into ontologies. However, the correlation between the resulting ontologies built automatically with plug-ins from relational databases and the database-to ontology mapping principles has been given little attention. This study reviews and applies two Protégé plug-ins, namely, DataMaster and OntoBase to automatically construct ontologies from a relational database. The resulting ontologies are further analysed to match their structures against the database-to-ontology mapping principles. A comparative analysis of the matching results reveals that OntoBase outperforms DataMaster in applying the database-to-ontology mapping principles for automatically converting relational databases into ontologies*
.

## 1. INTRODUCTION

Ontologies are an integral part of the growth and eventual realisation of the Semantic Web [1]. Ontology is an explicit specification of a conceptualisation that describes semantics of data [3]; it constitutes the backbone of Semantic Web applications [4], [6]. Due to the importance of ontologies in Semantic Web, researchers have proposed different methods and techniques to convert traditional relational databases into well-structured ontologies. In fact, relational databases remain an important source of data for many websites and applications [7],[8].

Ontology construction from a relational database used to be a manual and tedious process which relied solely on ontology editors and human experts [9]. Over the years, many tools and algorithms that enabled the automatic conversion of a relational database into ontology have been proposed. Examples of such tools and algorithms include: DB2OWL, R2O, D2RQ, Data

Semantic Preservation, DartGrid Semantic, Semantic Bridge, Automapper,XTR-RTO, RTAXON, Leaning Ontology from Relational Databases, Ontology Generator (RDB2On), and RDBToOnto amongst others [10], [11], [12], [13]. The World Wide Web Consortium (W3C) through their RDB2RDF Working Group are also developing a direct mapping standard that focuses on translating relational database into RDF (Resource Description Framework) ontology [2].

The problem with many of the abovementioned tools is that they are still at the prototype stage and are not yet available to the public. In fact, some of these tools are still under development and are not yet fully fledged products. Furthermore, these tools have not yet been applied on real world databases to ascertain their performance in the automatic conversion of relational databases into ontologies. Protégé is a widely used ontology editing platform which offers great extensibility and scalability [14]. Its extensibility is due to plug-ins developed by Semantic Web experts. A plug-in is a separately developed software module that adds more functionality to existing software. Examples of Protégé plug-ins include OntoLT [15], SIM-DLA [16], DataMaster [17], DataGenie[18], OntoBase [19] and RONTO [20]. OntoLT enables the extraction of ontology from text within Protégé [15].SIM-DLA is a Protégé plug-in that enables the comparison of ontology concepts and their meanings through the measurement of semantic similarities [16]. DataMaster, DataGenie, OntoBase and RONTO are Protégé plug-ins that deal with the conversion of relational databases into ontologies. However, the RONTO plug-in is still under development and is not yet available for use in the Semantic Web community [20]. Further, due to technical challenges such as unresolved errors and bugs [18], DataGenie functionalities were improved to create the DataMaster plug-in [17].

In light of the above, DataMaster and OntoBase are the only plug-ins for automatic conversion of relational databases into ontologies that are currently available for use in Protégé. However, their performances in accurately applying the database-to-ontology mapping principles [2], [9], [12], [21], [22], [23], [24] are still unreported. In fact, to the best of our knowledge, the database-to-ontology mapping performances of DataMaster and OntoBase plug-ins have not been reported in any previous study. This study aims at filling this gap in the Semantic Web literature. The DataMaster and OntoBase plug-ins are applied to automatically construct ontologies from a relational database. The resulting ontologies are further analysed to match their structures against the database-toontology mapping principles. A comparative analysis of the matching results reveals that OntoBase outperforms DataMaster in applying the database-to-ontology mapping principles.

The rest of the paper is organized as follows. Section 2 presents existing Semantic Web tools and algorithms for the conversion of relational databases into ontologies. The formal structure of a relational database as well as the database-to-ontology mapping principles are discussed in Section 3. Section 4 presents the experiments and results of the application of the database-to-ontology mapping rules/principles defined in Section 3. The paper ends with a conclusion and future work in Section 5.

## 2. REVIEW OF DATABASE-TO-ONTOLOGY MAPPING TOOLS AND ALGORITHMS

As mentioned earlier, various tools and algorithms have been developed to support the conversion of relational databases into ontologies. A relational database to ontology mapping tool called DB2OWL is presented in [21].DB2OWL enables the automatic generation of OWL ontologies from relational database schemas. However,DB2OWL is still at the prototype stage and not yet available for widespread utilisation in the Semantic Web community.

In [17], a Protégé plug-in, namely, DataMaster is presented. The DataMaster plug-in enables the import of schema structure and data from relational database into Protégé-OWL or Protégé-Frames ontology. OntoBase,another Protégé plug-in is presented in [19]. It utilises reverse-engineering to create ontology from a relational database schema [19]. Both DataMaster and OntoBase Protégé plug-ins are open source software that can be downloaded free of charge from the Internet. DataMaster comes as part of the Protégé ontology editor package while OntoBase can be downloaded separately [19].

In addition to DataMaster and OntoBase, other Protégé plug-ins, namely, RONTO and DataGenie are presented in [20] and [18], respectively. RONTO is described as a semiautomatic tool that enables schema matching between relational schemata and ontologies. However, RONTO was implemented as a prototype and is not yet available for use in the Semantic Web community. DataGenie is another Protégé plug-in which was developed to enable the import of a relational database into ontology. However, due to technical challenges such as unresolved errors and bugs [18], DataGenie functionalities were improved to create the DataMaster plug-in [17].

An algorithm called MARSON (Mapping between Relational Schemas and Ontologies) is presented in [1]. The algorithm establishes simple mappings between a relational database schema and ontology. Another algorithm,namely, RTAXON is presented in [25]. It enables the building of ontology from the schema definition and data stored in the database [25]. The RTAXON algorithm is implemented in a prototype application called RDBToOnto [13], [26]. Another prototype ontology generator, namely, RDB2on is presented in [12]. The RDB2on uses predefined transformation rules to automatically transform a relational database to OWL Ontology [12]. The database-to-ontology mapping rules/principles are discussed in the next section.

## 3. RELATIONAL DATABASE TO ONTOLOGY MAPPING

In this section, relational database, ontology and database-to-ontology mapping rules/principles are defined to set  the conceptual background of the study.

### 3.1 Relational Database

A relational database is a data model which includes sets of relationships, attributes, and basic types [24]. A relational database could be represented in the form of a relational database schema [25]. The relational database schema defines the structure of the database [26] and consists of the following main elements [16], [21], [27], [24],[28]:

• Relation - database table with a set of columns, rows and constraints.
• Attribute - column of a database table.
• Tuple - record or row of a database table.
• Domain - data type of a column of a database table. This is the type of values that a column can have e.g.
Integer values etc.
• Primary Key - a constraint placed on a column to maintain entity integrity in the table. A primary key maintains unique rows in the table.
• Foreign Key - a constraint placed on a column to maintain referential integrity. A foreign key maintains relationships among database tables.

A relational database can have different types of relationships between its tables. The relationships are maintained by the use of foreign keys. Let's consider two related tables T1 and T2 with sets of rows R1 and R2, respectively.The possible relationships between the tables of the relational database are as follows:

• One to One relationship - one row $r1i \in R1$ ($1 \le i \le n$) in T1 corresponds to only one row $r2j \in R2$ ($1 \le j \le m$) in T2, where n and m are the numbers of rows in T1 and T2, respectively, i.e., only one row in T1 corresponds to only one row in T2.
• One to Many relationship - each row $r1i \in R1$ ($1 \le i \le n$) in T1 corresponds to $s2j \in R2$ ($1 \le j \le m$) in T2, where $s2j = \{r2k, 1 \le k \le m\}$ is a set of rows in T2, n and m, the numbers of rows in T1 and T2, respectively. This means that one row in T1 can have many corresponding rows in T2. In this relationship, a primary key in T1 will be a foreign key in T2.
• Many to Many relationships - a set a rows $s1i \in R1$ ($1 \le i \le n$) in T1 corresponds to a set of rows $s2j \in R2$ ($1 \le j \le m$) in T2, where n and m are the number of rows in T1 and T2,respectively, i.e., many rows in T1 corresponds to many rows in T2. These relationships are normally resolved by a use of bridge tables.

## 3.2 Definition of Ontology

Ontology is a knowledge base system representing the common and shared vocabularies/concepts within a specific domain as well as the relationships between them [9], [16], [27]. Typical ontology elements are concepts,relationships/properties, axioms and instances [24], [28]. A concept is the basic component of ontology. The relationships/properties between concepts define how concepts are semantically related to each other in the ontology. Axioms are the statements in the ontology, i.e., the logical combinations of concepts and properties. The instances are the occurrences/values of concepts or properties in the ontology. The popular languages for the formal representation of ontology are RDF and Web Ontology Language (OWL). However,OWL is preferred over RDF [9], [30], due to the weak expressive power of the RDF language [29],[30].It is also said to be the most advanced ontology representation language [31]. The common keywords of the OWL language for representing ontology elements are defined below [20], [29], [32], [24], [27], [31]:
1. *Class:*

It represents a concept of an ontology in OWL [9, 30]. An example of OWL representation of a class named PropertyType is given in the line of code below.

```
<owl:Class rdf:ID = "#PropertyType" />
```

2. *Object Property:*

This OWL construct defines relationships between ontology classes [24]. Object Properties are defined using domains and ranges which are the classes that are in relation with one another [16]. The following code presents an OWL Object Property named PropertyTypeIDInstance. The domain of the PropertyTypeIDInstance Object Property is the PropertyService class and its range the PropertyType class, i.e., PropertyService and PropertyType are in a relation with one another.

```
<owl:ObjectProperty rdf:ID="#PropertyTypeIDInstance"/>
<rdf:type rdf:resource="#functional property"/>
<rdfs:domain rdf:resource="#PropertyService"/>
<rdfs:range rdf:resource="#PropertyType"/>
</owl:ObjectProperty>
```

3. *Datatype Property:*

It represents the attributes of ontology classes in OWL [20]. Datatype Properties are also defined using domains and ranges; here, the domain represents a class that the property belongs to and range represents the type and limit of data that the property can store [16]. An example of OWL Datatype Property named Description is given in the code below. The domain of the Datatype Property is the PropertyType class and the range is String. The range indicated that Description Datatype Property represents string values.

```
<owl: DatatypeProperty rdf:ID="#Description"/>
<rdfs: domain rdf:resource="#PropertyType"/>
<rdfs:range rdf:resource="XMLSchema#string"/>
</owl:DataTypeProperty/>
```

4. *Individual:*

 It is an instance of a class or property. An example of an Individual named PropertyTypeInstance is given in the OWL code below. This is an instance of the PropertyType class.

```
<owl:PropertyType rdf:ID="#PropertyTypeInstance"/>
<owl: PropertyTypeID rdf: datatype="&xsd; int">1< owl:PropertyTypeID/>
<owl:Descripion rdf:datatype="XMLSchema#string">Residential
<owl:Description/>
<owl:Ratable rdf:datatype="XMLSchema#string">Yes <owl:Ratable/>
```
Class, Object Property, and Datatype Property are the main OWL elements as they represent ontology concepts,relationships between the concepts and attributes of the concepts. Classes and properties are components upon which the ontology hierarchy is built [29]. In the OWL hierarchy, owl:Thing is the base class and any other class in the ontology inherits from it [20]. The next Subsection presents a review of existing mapping rules that govern the conversion of a relational database into O WL ontology.

| Relations | Primary Key | Foreign Key |
|---|---|---|
| PropertyType(PropertyTypeID, Description) | PropertyTypeID | None |
| PropertyService(PropertyServiceID, **ServiceID, PropertyTypeID**) | PropertyServiceID | **ServiceID** refers to ServiceID in Service |
| | | **PropertyTypeID** refers to PropertyTypeID in PropertyType |
| Service(ServiceID, Description, Type) | ServiceID | None |
| Customer(CustomerID, Name, ID, RegistrationNo, PhysicalAddress, PostalAddress, Telephone, CellPhone, Email) | CustomerID | None |
| Query(QueryID, **CustomerID**, Status, Type, DateEntered, DateClosed, Details, **AttendedBy**) | QueryID | **CustomerID** refers to CustomerID in Customer |
| | | **AttendedBy** refers to EmployeeID in Employee |

Table 1: Sample Relational Database Schema

## 3.3 Relational Database to Ontology Mapping Rules/Principles

The process of converting a relational database into ontology follows certain mapping rules/principles [16], [20],[22], [24], [27], [29], [32], [33]. Mapping rules define how relational database components including Tables,Columns, Foreign Keys, etc., can be converted into ontology components such as Classes, Properties, Instances,etc. In this Subsection, existing mapping rules are discussed using a sample relational database schema in Table 1.The mapping rules used to convert the database tables in Table 1 into OWL ontology constructs are presented below.

1. *Rule 1*

*Mapping of Tables to OWL Classes:* Each table in the relational database is mapped into ontology OWL class with similar name except for bridging bridge tables that are used to resolve many-to-many relationships [20], [16], [32], [24]. On that note, only all the four tables in Table 1 are mapped to OWL classes as in the sample code below.

```
<owl:Class rdf:ID = "#PropertyType" />
<owl:Class rdf:ID = "#Service" />
<owl:Class rdf:ID = "#Customer" />
<owl:Class rdf:ID = "#Query" />
<owl:Class rdf:ID = "PropertyService" />
```

The PropertyService table (Table 1) was also converted to an OWL class because it couldn't be recognised as a bridge table even though it is used to resolve a many-to-many relationship between PropertyType and Service database tables. This is because it has a separate PropertyServiceID Primary Key in addition to the two foreign keys (ServiceID and PropertyTypeID). Rule 2 underneath elaborates more on handling of bridge tables.

2. *Rule 2*

*Handling of Bridge Tables:* Bridge tables are not mapped into separate OWL classes. This rule applies to properly constructed bridge tables which have foreign keys from the tables participating in a many-tomany relationship as its main primary keys. Even though there is no separate class, many-to-many relationships are still represented by Object Properties in the ontology [33]. More on Object Properties is covered in Rule 6 and 7 underneath.

3. *Rule 3*

*Mapping of Referential Integrity Relationships to Inheritance Hierarchy:* OWL Classes are arranged in a hierarchy based on the relationships in the database. In a relationship between two tables, a table that has a foreign key will be mapped into a sub-class of the main class obtained from a table with a corresponding primary key. For example, from the classes created in Rule 1 above, Query will be a sub class of Customer because of a relationship between Query and Customer tables. Query table has a CustomerID foreign key to symbolise its
dependence on the Customer table. An example of OWL code is depicted below:

*<owl:Class rdf:ID = "#Query">*
*<rdfs:subClassOf rdf:resource="#Customer"*
*<owl:Class />*

4. *Rule 4*

*Mapping of Non-Referential Integrity Columns into Datatype Properties:* All columns in the relational database are mapped into Datatype Properties, except all the foreign keys which maintain referential integrity in the database [20], [24], [32], [33]. For instance, the Query class obtained in Rule 1 will have QueryID,Status, Type, DateEntered, DateClosed, and Details as Datatype Properties. CustomerID and AttendedBy are excluded from Datatype Properties list. The basic OWL code of the Datatype Property named Details in the Query class is provided below.

*<owl: DatatypeProperty rdf:ID = "#Details" />*
*</owl:DatatypeProperty/>*

5. *Rule 5*

*Representation of Datatype Property host class as Domain and Data Type as Range:* A Datatype Property includes domain and range which represents the host class and the type of data that will be represented,respectively [20], [24], [32], [33]. The code below shows the *Query* class as the domain of the *Details* Datatype Property, whereas, its range is the string datatype, i.e., the Details Datatype Property will represent string values.

```
<owl: DatatypeProperty rdf:ID = "#Details" />
<rdfs:domain rdf:resource = "#Query" />
<rdfs:range rdf:resource = "XMLSchema#string" />
</owl:DatatypeProperty/>
```

## 6. *Rule 6*

*Mapping of Relationships represented by referential integrity columns into Object Properties:* All relationships that are expressed with foreign keys in a relational database are mapped into OWL Object Properties [16], [20], [29], [32]. Two Object Properties are created for one-to-many or a many-to-many relationship, one for the relationship and one for its inverse. For instance, the Query and Customer classes obtained in Rule 1 would produce two Object Properties which are represented by a CustomerID Functional Property within the Query class and a CustomerID Inverse Functional Property within the Customer class. This is because the Query class was derived from a Query table with a foreign key that points to a primary key in the Customer table. An OWL code for the Object Properties between the Query and Customer classes is given below:

```
<owl:ObjectProperty rdf:ID = "#CustomerIDInstance"/>
<rdf:type rdf:resource = "#functional property" />
</owl:ObjectProperty />
<owl:ObjectProperty rdf:ID = "#CustomerIDInstance"/>
<rdf:type rdf:resource = "#inversefunctional property" />
</owl:ObjectProperty />
```

## 7. *Rule 7*

*Representation of Object Property host classes as domain and range:* An Object Property includes domain and range which represent the two classes in relation with one another. The domain is a class with a functional property while a range is a class with an inverse functional property [16], [20], [29], [32]. From the code shown below, the domain of the Object Property *CustomerIDInstance* is the Query class, whereas, its range is the Customer class. This Object Property defines the semantic relationship between the Query and Customer classes in the ontology.

```
<owl:ObjectProperty rdf:ID = "#CustomerIDInstance"/>
<rdf:type rdf:resource = "#functional property" />
<rdfs:domain rdf:resource = "#Query" />
<rdfs:range rdf:resource = "#Customer" />
</owl:ObjectProperty />
```

8. *Rule 8*

*Mapping of Tuples to Individuals:* All database table records are mapped to individuals in ontology [20], [29], [24], [32]. For instance, if the Service table from Table 1 had two rows of data, those rows will be mapped to OWL individuals as in the code below:

```
<owl:Service rdf:ID="#ServiceInstance"/>
<owl: ServiceID rdf: datatype="&xsd;int">1
</ owl: ServiceID>
<owl: Description rdf: atatype="XMLSchema#string">Electricity
<owl: Description/>
<owl: Type rdf: datatype="XMLSchema#string">Consumable <owl: Type/>
<owl: Service rdf: ID="#ServiceInstance2"/>
<owl: ServiceID rdf: datatype="&xsd;int">2
</ owl:ServiceID>
<owl: Description rdf: datatype="XMLSchema#string">Refuse Removal </owl: Description>
<owl: Type rdf: datatype="XMLSchema#string">Basic
</owl: Type>
```

9. *Rule 9*

*Mapping of Column Constraints into Property Cardinalities:* Database column constraints e.g. NULL and NOT NULL are mapped into Ontology Property Cardinalities [22], [29]. Cardinalities are there to further specify and place restrictions on ontology properties [29]. For example, let us say the Query table in Table 1 has a QueryID column which is declared as NOT NULL and a Type column which is NULL. This will lead to the following cardinalities in the ontology:

```
<owl:Restriction>
<owl:onProperty rdf:resource="#QueryID"/>
<owl:minCardinality>1< owl:minCardinality/>
<owl:maxCardinality>0< owl:maxCardinality/>
<owl:Restriction/>
<owl:Restriction>
<owl:onProperty rdf:resource="#Type"/>
<owl:minCardinality>0< owl:minCardinality/>
<owl:maxCardinality>1< owl:maxCardinality/>
<owl:Restriction/>
```

# 4. EXPERIMENTS

## 4.1. Dataset

An Oracle database developed in [32] was modified and used as the input database in the experiments in this study.The database was built from the Municipality Information System for service delivery in the authors' country [32].The main tables of the database are Account, AccountService, Arrangement, Arrears, Customer, Employee,Manager, Payment, Property, PropertyService, PropertyType, Query, Service, Tariff, CustomerType, Penalty,Rebate, ValuationRoll and ValuationRollType. This study does not expand on the process used to build the entity relationship diagram (ERD) from which the database was developed. Interested readers may refer to the work in [32] for further information. Figure 1 presents the Municipality database schema in a tabular format including all the tables, columns and relationships that exist in the database.

| Relation | Primary Key | Foreign Key |
|---|---|---|
| Account (AccountID, CustomerID, PropertyID, DateOpened, Balance, Status) | AccountID | **CustomerID** refers to CustomerID in Customer |
| | | **PropertyID** refers to PropertyID in Property |
| AccountService (AccountServiceID, AccountID, PropertyServiceID, Date, Status, Consumption, Reading) | AccountServiceID | **AccountID** refers to AccountID in Account |
| | | **PropertyServiceID** refers to PropertyServiceID in PropertyService |
| Arrangement (ArrangementID, AccountID, Date, Balance, Instalment, Status) | ArrangementID | **AccountID** refers to AccountID in Account |
| Arrears (ArrearsID, AccountID, Date, Age, Amount) | ArrearsID | **AccountID** refers to AccountID in Account |
| CustomerType (CustomerTypeID, Description) | CustomerTypeID | None |
| Customer (CustomerID, CustomerTypeID, Name, ID-RegistrationNo, PhysicalAddress, PostalAddress, Telephone, Cellphone, Email) | CustomerID | **CustomerTypeID** refers to CustomerTypeID in CustomerType |
| Employee (EmployeeID, ManagerID, Name, Surname, Telephone, Cellphone, Email) | EmployeeID | **ManagerID** refers to ManagerID in Manager |
| Manager (ManagerID, Name, Surname, Telephone, Cellphone, Email) | ManagerID | None |
| Payment (PaymentID, AccountID, Date, Method, Amount) | PaymentID | **AccountID** refers to AccountID in Account |
| Penalty (PenaltyID, AccountID, Date, Age, Amount, Interest, Description) | PenaltyID | **AccountID** refers to AccountID in Account |
| Property (PropertyID, PropertyTypeID, StandNo, StandAddress, RegisteredOwner, OwnerAddress, DateRegistered, MarketValue, Status) | PropertyID | **PropertyTypeID** refers to PropertyTypeID in PropertyType |
| PropertyService (PropertyServiceID, ServiceID, PropertyTypeID) | PropertyServiceID | **ServiceID** refers to ServiceID in Service |
| | | **PropertyTypeID** refers to PropertyTypeID in PropertyType |
| PropertyType (PropertyTypeID, Description) | PropertyTypeID | None |
| Query (QueryID, CustomerID, Status, Type, DateEntered, DateClosed, Details, AttendedBy) | QueryID | **CustomerID** refers to CustomerID in Customer |
| | | **AttendedBy** refers to EmployeeID in Employee |
| Rebate (RebateID, AccountID, Amount, Description) | RebateID | **AccountID** refers to AccountID in Account |
| Service (ServiceID, Description, Type) | ServiceID | None |
| Tariff (TariffID, ServiceID, PropertyTypeID, StartDate, EndDate, Price) | TariffID | **ServiceID** refers to ServiceID in Service |
| | | **PropertyTypeID** refers to PropertyTypeID in PropertyType |
| ValuationRoll (ValuationRollID, ValuationRollTypeID, PropertyID, LastValuer, StartDate, EndDate, MarketValue) | ValuationRollID | **ValuationRollTypeID** refers to ValuationRollTypeID in ValuationRollType |
| | | **PropertyTypeID** refers to PropertyTypeID in PropertyType |
| | | **LastValuer** refers to EmployeeID in Employee |
| ValuationRollType (ValuationRollTypeID, Description, Duration) | ValuationRollTypeID | None |

Figure 1: Oracle Database Schema for Municipality Information System [32]

The screenshot in Figure 2 shows a view of the Municipality database in Figure 1 created in Oracle 11g. Oracle was chosen as Relational Database Management System (RDBMS) because it is highly recommended for Semantic Web development [12], [21]. The database in Figure 2 was initially migrated into Oracle from Microsoft SQL Server [32].

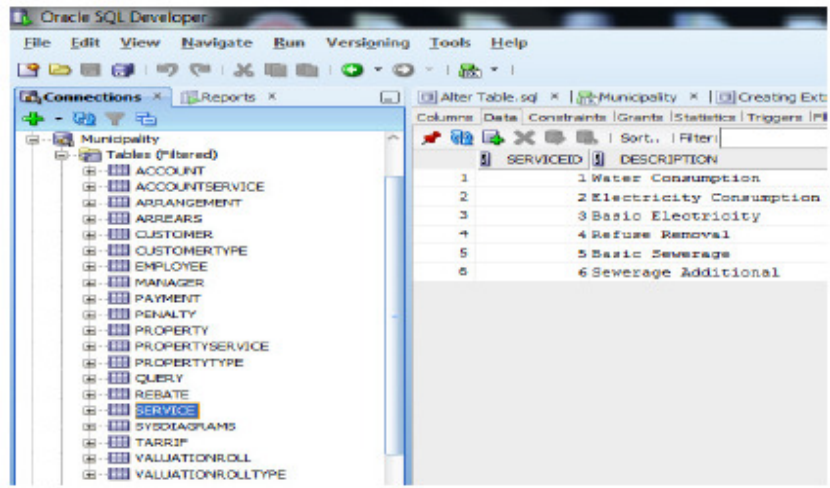

Figure 2: Screenshot of the Oracle Database

It can be noticed in Figure 2 that the database has an extra Sysdiagrams table which wasn't included in Figure 1.The Sysdiagrams table was automatically created by Oracle during the migration from Microsoft SQL Server. This table is not a critical part of the database. Furthermore, the database was loaded with test data before the experiments, i.e, four rows in the *Manager* table and six rows in the *Service* table (Figure 2)

## 4.2. Computer and Software Environment

Experiments were carried out on a Dual Core 32 bit Notebook with 2 GB of RAM and a Windows 7 Operating System. It is important to recall that, Oracle 11g Express Edition was used as RDBMS. Two plug-ins, namely,DataMaster [17] and OntoBase [19] were used to automatically construct ontologies from the Oracle database in Protégé version 3.5. Both plug-ins utilize the Oracle JDBC driver to establish a connection to the Oracle database.The graphical representation of the output ontologies from DataMaster and OntoBase was done using virtualization plug-ins including OntoGraf [34] and OWLViz [33]. A Semantic Web tool that generates a structured documentation of ontology, namely, Parrot [35] was used to display and analyse the structure of the output ontologies codes from DataMaster and OntoBase.

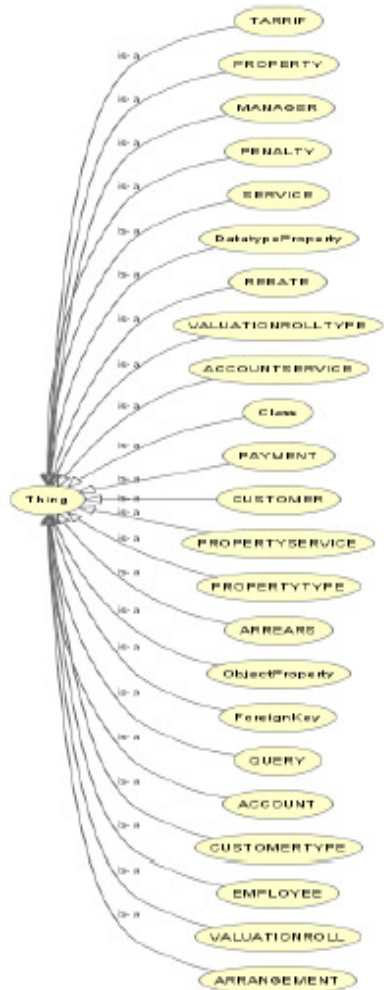

Figure 3: Inheritance Structure of Ontology Constructed with DataMaster Plug-in via OWLViz

## 4.3 Experimental Results

Figure 3 shows the classes of the OWL ontology constructed from the Oracle database (Figure 1 and 2) with the DataMaster plug-in. The graphical representation of classes in Figure 3 was obtained with the OWLViz virtualisation plug-in. The complete graph of the resulting ontology is shown in Figure 4; this graph was generated with the OntoGraf [34] virtualisation plug-in. Figure 4 shows all the classes of the ontology constructed with the DataMaster plug-in and the relationships between them.

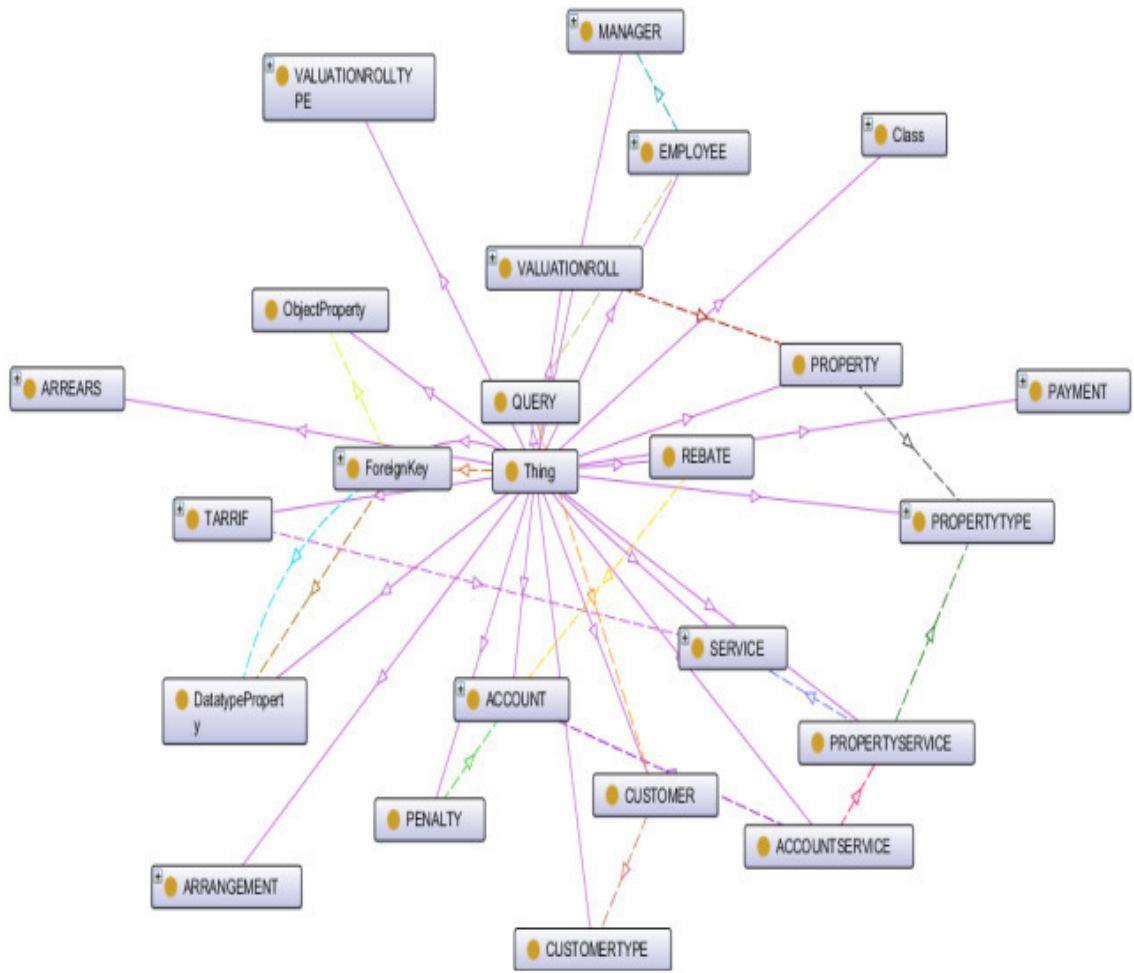

Figure 4: Ontology Constructed with DataMaster Plug-in

Similarly, OntoBase plug-in was used to construct OWL ontology from the Oracle database (Figure 1 and 2).Figure 5 shows a screenshot of the ontology constructed with OntoBase within Protégé. The ontology in Figure 5 was further represented graphically with the OntoGraf visualization plug-in as in Figure 6. In Figure 6, all classes of the resulting ontology and the relationships between them are shown.

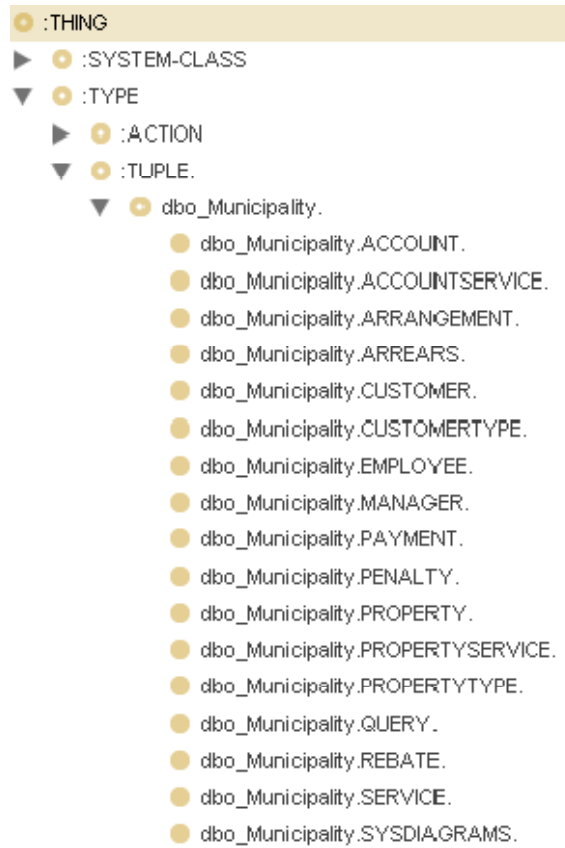

Figure 5: Screenshot of Ontology Constructed with OntoBase Plug-in

As mentioned earlier, the OWL codes of the ontologies constructed with both DataMaster and OntoBase plug-ins were further analysed using the Parrot [35] ontology documentation software. Parrot displayed the structure of the resulting OWL ontologies as well as useful comments that explained the OWL constructs (classes, Datatype Properties, Object Properties, etc.) within the ontologies.

Figure 7 shows the mapping results of the Oracle database (Figure 1 and 2) into ontology (Figure 3 and 4) with the DataMaster plug-in. The results in Figure 7 (a and b) shows that all tables were successfully mapped to ontology classes (Figure 3 and 4) including the PropertyService table which is used to resolve the many-to-many relationship. In fact, the PropertyService table has a Primary Key and cannot be treated as a bridge table. In addition to the classes mapped from the relational database tables, four other classes were created.

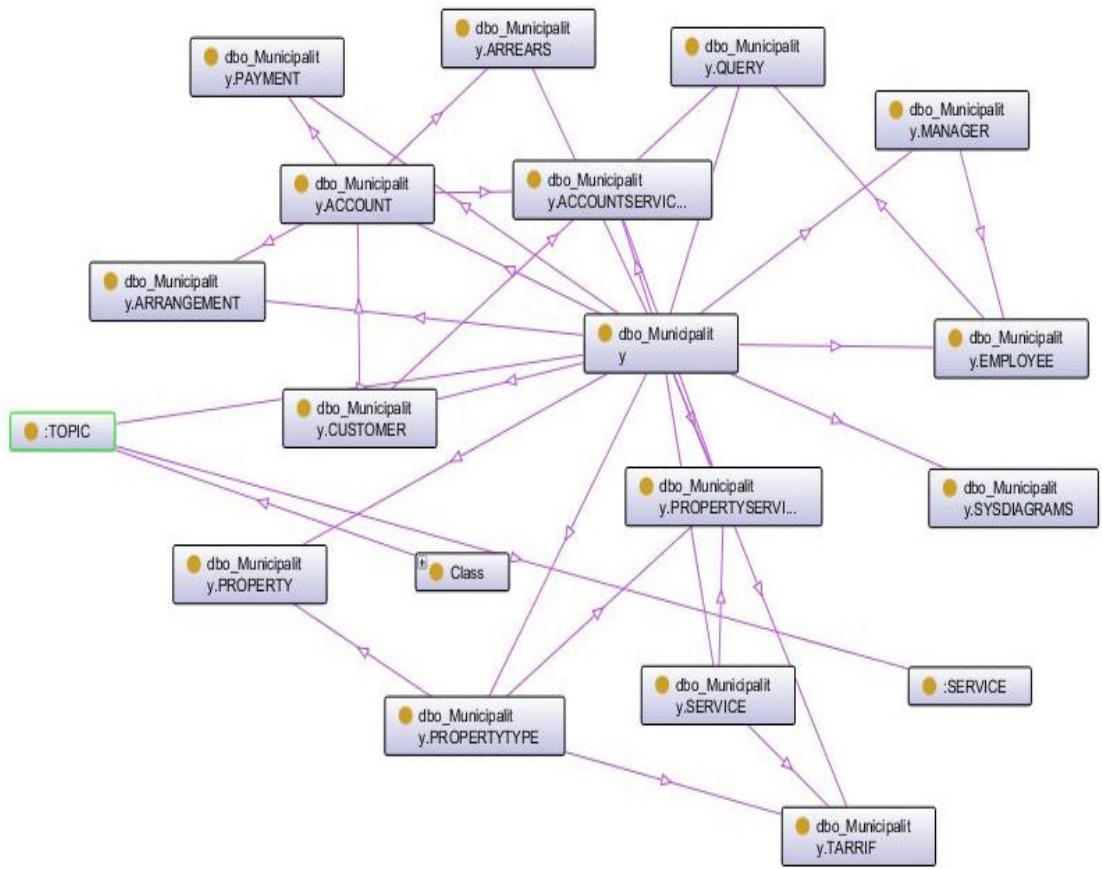

Figure 6: Part of Ontology Constructed with OntoBase

With regards to Datatype Properties, Figure 7 (a, b) shows that all 105 columns were mapped into Datatype Properties irrespective of whether they were foreign keys or not. DataMaster also added 7 extra Datatype Properties. This finding reveals a slight deviation from the mapping principles in Subsection 3.3. The results in Figure 7(b) shows that all 21 foreign keys in the input Oracle database were successfully mapped to Object Properties with the addition of 4 more Object Properties. A deviation here is that duplicate Object Properties were not created to represent inverse functional properties as stated in the mapping principles in Subsection 3.3.

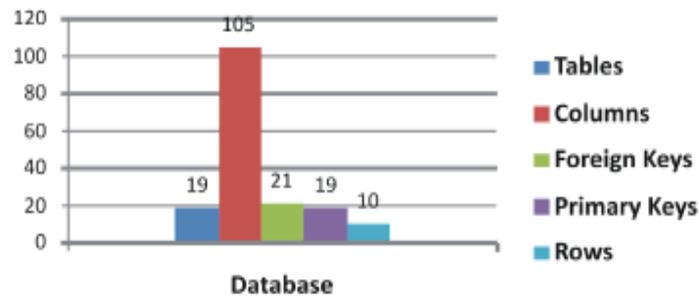

(a)

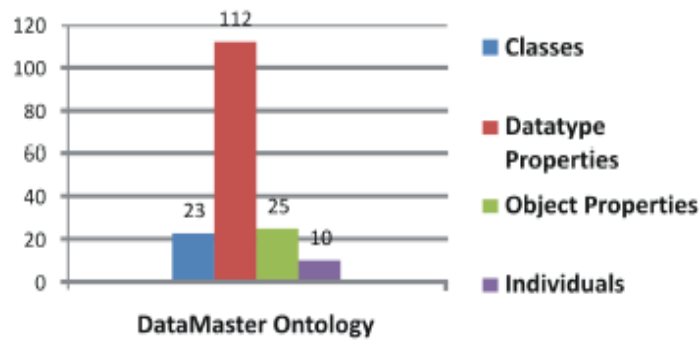

Figure 7: (a) Database components (b) DataMaster Ontology Components

The results in Figure 7(b) shows that all 10 test rows in the input Oracle database were successfully mapped to Individuals. Lastly, the DataMaster plug-in did not produce any cardinality although there were NULL and NOT NULL columns in the database. However, it handled bridge tables according to the mapping principles.Further more, although DataMaster generated an inheritance hierarchy as in Figure 3, it did not comply with the mapping principles in Subsection 3.3. Overall, the results in Figure 7 show that, according to database-to-ontology mapping principles, the ontology constructed with the DataMaster plug-in captured most of the features of the input database even though there were slight deviations from the mapping principles.

Similarly, Figure 8 shows the mapping results of the Oracle database (Figure 1 and 2) into ontology (Figure 5 and 6) with the OntoBase plug-in. Figure 8 (a and b) shows that all tables in the database were successfully mapped to ontology classes with an addition of 1 class. The PropertyService table was also converted into a class. In fact, the PropertyService table has a Primary Key and could not be traited as a bridge table.

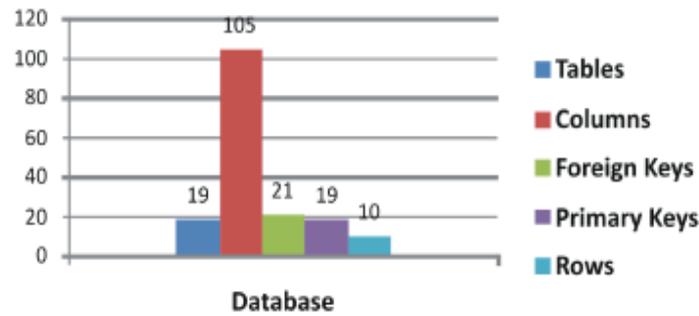

(a)

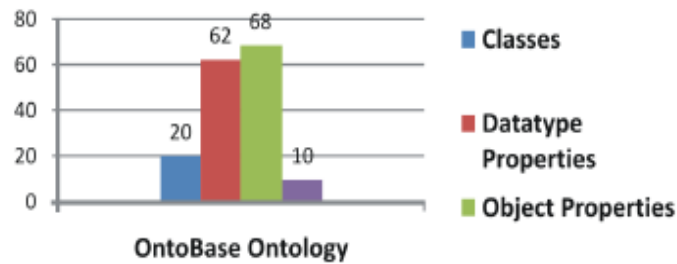

(b)

Figure 8: (a) Database components (b) OntoBase Ontology Components

Results in Figure 8 (a and b) shows that all columns of the database were mapped into Datatype Properties except for foreign keys. In fact, the OntoBase plug-in produced fewer Datatype Properties; this proves that foreign key columns were indeed excluded. This is a major conformance with the mapping principles in Subsection 3.3.

With regard to Object Properties, the results in Figure 8 (b) show that all foreign keys were successfully mapped to Object Properties with all the necessary duplicates due to one-to-many and many-to-many relationships. In Figure 8 (b), it is shown that all 10 test rows in the input Oracle database were successfully mapped to Individuals.Lastly, similar to DataMaster, the OntoBase plug-in did not produce any cardinality although there were NULL and NOT NULL columns in the database. However, as DataMaster, it also handled bridge tables according to the mapping principles. Furthermore, different from DataMaster, OntoBase generated an inheritance hierarchy that complies with the mapping principles in Section 3.3. Overall the results in Figure 8 reveal that the structure of the ontology obtained with the OntoBase plug-in has few deviations and does capture accurately the features of the input database according to the database-to-ontology mapping principles in Subsection 3.3.

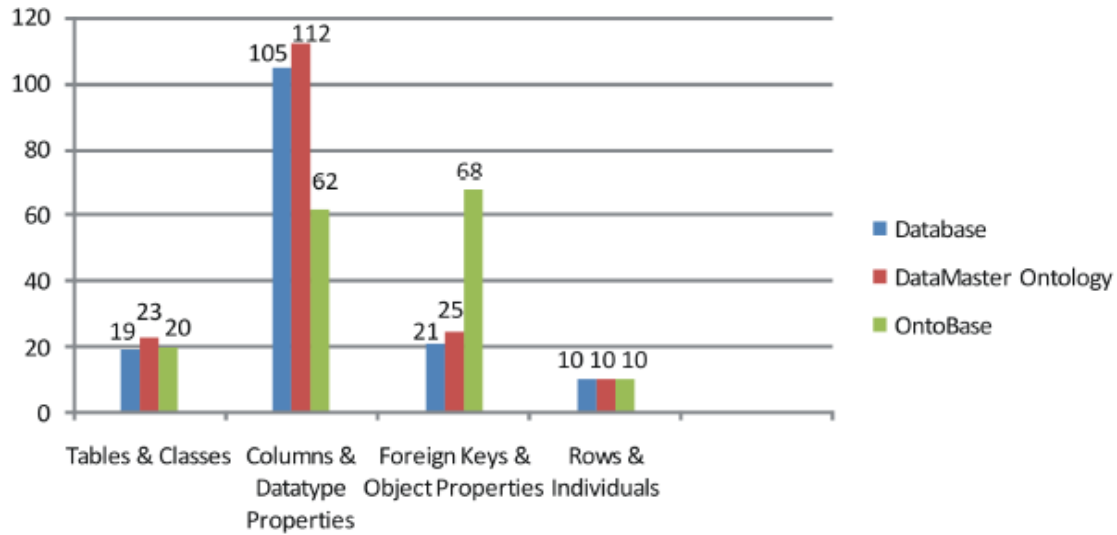

Figure 9: Chart of Comparison of Performances of DataMaster and OntoBase Plug-ins

Figure 7 and 8 presented separate results for DataMaster and OntoBase plug-ins. In Figure 9, all the results are tallied and the mapping performances of both plug-ins are compared. It is shown in the left block of Figure 9 that 19 database tables (left bar) are mapped into 23 ontology classes in DataMaster (middle bar) and 20 ontology classes in OntoBase (right bar). In the second left block of Figure 9, 105 database columns (left bar) are mapped into 112 Datatype Properties in DataMaster (middle bar) and 62 Datatype Properties in OntoBase (right bar). The second right block of Figure 9 depicts 21 database foreign keys (left bar) that are mapped into 25 Object Properties in DataMaster (middle bar) and 68 Object Properties in OntoBase (right bar). These results reveal that DataMaster has more deviations from the mapping principles as far as producing an accurate Ontology from the relational database is concerned. OntoBase on the other hand conformed with the mapping principles in Subsection 3.3; this conformance is witnessed in the low number of Datatype Properties (62) in the resulting ontology compared to the number of columns (105) in the input database as well as the high number of Object Properties (68) compared to the number of foreign keys (21) in the input database.

## 5. CONCLUSION

In this study, ontologies were automatically constructed from an Oracle relational database with two Protégé plugins,namely, DataMaster and OntoBase. The semantic structures of the resulting ontologies were analysed by means of two visualization plug-ins including OntoGraf and OWLViz as well as an ontology documentation software, namely, Parrot. The performances of the plug-ins were further measured based on the database-toontology mapping rules/principles. The results revealed that both tools reasonably convert a relational database to ontology with slight deviations from the database-to-ontology mapping principles. The results of the studies provide interesting insights on the performance and accuracy of Protégé plug-ins in converting relational databases into ontologies; this may be useful to developers who are developing Sematic Web applications that interface legacy relational databases of organizations.The future direction of the research would be to repeat the experiments with larger relational databases and measure the scalabilities of DataMaster and OntoBase plug-ins. Another addition is to expand the study with Semantic Web tools other than Protégé.